\documentclass[runningheads]{llncs}

\usepackage{array} % for better arrays (eg matrices) in maths
\usepackage{amsmath} % Math environment
\usepackage{url} % url in bibliography
\usepackage{graphicx} % To add images
\usepackage{multirow} % To merge rows in a table

\newcommand{\bigcb}[1]{\hbox{$\left\{\vbox to#1cm{}\right.\n@space$}}

\begin{document}
\title{User Experience Design for E-Voting:\\ \large{How mental models align with security mechanisms}}
\author{Marie-Laure Zollinger \inst{1} \and Verena Distler \inst{1} \and Peter B. R\o nne \inst{1} \and Peter Y. A. Ryan \inst{1} \and Carine Lallemand \inst{2} \and Vincent Koenig \inst{1} } 
\institute{University of Luxembourg, Esch-sur-Alzette, Luxembourg
\\\email{\{marie-laure.zollinger,verena.distler,peter.roenne,peter.ryan, vincent.koenig\}@uni.lu}
\and
 Eindhoven University of Technology, Netherlands
\\\email{c.e.lallemand@tue.nl}} 
\authorrunning{M. Zollinger et al.}
\titlerunning{User Experience Design for E-Voting}

\maketitle

\begin{abstract}
This paper presents a mobile application for vote-casting and vote-verification
based on the Selene e-voting protocol and explains how it was developed and implemented using the User Experience Design process. The resulting interface was tested with 38 participants, and user experience data was collected via questionnaires and semi-structured interviews on user experience and perceived security. Results concerning the impact of displaying security mechanisms on UX were presented in a complementary paper \cite{verena}.
Here we expand on this analysis by studying the mental models revealed during the interviews and compare them with theoretical security notions. Finally, we propose a list of improvements for designs of future voting protocols.
\end{abstract}

\section{Introduction} \label{sec:intro}

Voting protocols are carefully designed to satisfy certain security properties, most importantly Privacy and End-to-End (E2E) Verifiability. Some notable privacy properties are ballot-secrecy, receipt-freeness and coercion-resistance. 
E2E verifiability is usually separated into the votes being cast-as-intended, recorded-as-cast, and tallied-as-recorded.

E2E-verifiable schemes often require voters to handle encrypted ballots \cite{DBLP:conf/uss/Adida08,DBLP:journals/corr/abs-1211-1904,DBLP:conf/wpes/JuelsCJ05}. The Selene e-voting protocol \cite{DBLP:conf/fc/RyanRI16} has been designed in order to hide the cryptographic operations from the voter. Instead, each voter is assigned a private tracking number, which lets them verify that their vote has been correctly included in the tally. In the setup phase, a unique tracker number is secretly associated with each voter and cryptographically committed to the bulletin board. At the end of the election, the votes are posted on a public bulletin board along with the associated tracking numbers. To avoid coercion, the voters are notified of their tracking number only after the vote/tracker pairs have been published. This gives coerced voters the opportunity to identify a tracker that points to the coercer's candidate that they can then claim is theirs.  The hypothesis is that this mechanism is more intuitive, transparent and easy-to-use than the usual E2E verifiability:  where voters should check the encryption of their vote and then presence of this encryption of their vote on the bulletin board.

User tests on voting protocols have shown that schemes that provide security often have usability issues \cite{DBLP:conf/uss/AcemyanKBW14,DBLP:conf/chi/MarkyKRV18,DBLP:journals/iacr/KulykV18}. According to \cite{iso}, \emph{usability} measures the effectiveness, efficiency, and satisfaction of a software in a specified context of use. Effectiveness is the accuracy and completeness with which the users achieve their goals. Efficiency represents the resources expended for effectiveness. Satisfaction is defined by the comfort and acceptability of use. In the papers \cite{DBLP:conf/uss/AcemyanKBW14,DBLP:conf/chi/MarkyKRV18,DBLP:journals/iacr/KulykV18}, the effectiveness of vote casting, that is the ability to cast successfully a vote, has been at most $81.25\%$ \cite{DBLP:conf/chi/MarkyKRV18}. In addition, the meaning of the verification phase is not always well understood, which can lead to voters not performing the verification task or unintentionally aborting the task. 
Ensuring system usability is further complicated by the fact that elections occur rarely and voters are expected to understand and use a system they are not familiar with.

\emph{User Experience} is defined as ``a person's perceptions and responses that result from the use or anticipated use of a product, system or service'' \cite{isoux}. It considers emotions, psychological needs and temporal aspects of the interaction between the system and the user, and can measure a person's perceptions of system qualities such as attractiveness, ease of use and novelty, in addition to usability aspects.
To improve the user experience, user-centred methodologies have been developed in order to include the final users in the development of a product \cite{Norman:1986:UCS:576915,normaneveryday,cooper2014face,LallemandCarine2018MddU}. We will describe the process in detail in section \ref{sec:selene}.

In this paper we present two main contributions, the first is the development of a prototype interface for smartphones for the Selene e-voting protocol, following a user-centered design process \cite{Norman:1986:UCS:576915,normaneveryday,cooper2014face} called User Experience Design (UXD) Process \cite{LallemandCarine2018MddU}.
We will discuss the impact of our implementation choices on the initial protocol. 
Then we did a user study on which the primary goal of the interviews was to retrieve insights and interpretation of behavioural data and to complement and triangulate data from the questionnaires, results can be found in \cite{verena}.
Our second contribution is to study the gaps between voting research and users expectations for a voting system, by exploring the mental models of voters for Privacy and Verifiability expressed in the semi-structured interviews. We define mental models as the concepts in people's mind that represent their understanding of how things work \cite{normaneveryday}. This paper describes the first application of the UXD method for app development in the e-voting context and evaluates on its use.

The paper is organized as follows: Section \ref{sec:selene} describes the Selene mechanism and details the development of the mobile application following the user-oriented process. Section \ref{sec:protocol} describes the steps of the user tests that have been done for this study. Section \ref{sec:mm} provides an analysis of participants' interviews and describes the mental models for Privacy and Verifiability. Finally section \ref{sec:discussion} discusses the implementation, the mental models found and the limitations of the study. We conclude in section \ref{sec:ccl} by suggesting design improvements and discuss future work. 

\subsubsection{Related Work} %\label{sec:rw}

The study of mental models is useful to align the system design with the users' expectation of a system, reducing the possible interaction errors that could lead to additional security (or safety) issues. The subject has received little attention in voting, we relate our work to the few publications here and discuss them in detail in section \ref{sec:discussion}.

Mental models of verifiability in postal voting and paper voting have been explored by Olembo et al. through a survey conducted in Germany \cite{DBLP:conf/voteid/OlemboBV13}. They suggested breaches in the procedures that could lead to integrity issues and asked participants about different aspects of verifiability. 
Our approach is different as we do not mention possible security issues in our interviews but we have let the participants express themselves based on the experience of voting with our application (see section \ref{sec:protocol}).

Another paper from Acemyan et al \cite{DBLP:conf/uss/AcemyanKBW14} analyzed mental models for three voting schemes which are Helios, Pr\^et \`a Voter and Scantegrity II. The experiment aimed to study the participants' mental models through drawings and interviews after using each of the voting systems. The analysis of participants' feedback showed that many participants did not see the E2E-verifiable schemes as being more secure than a standard paper-based voting method. The authors also highlighted that 
participants tend to focus more on the voting phase, we noticed a lack of understanding for verifiability as described in section \ref{sec:mm}.

Human factors in security were highlighted by Kulyk and Volkamer in \cite{DBLP:journals/iacr/KulykV18}. They extract five concepts including concern and self-efficacy, as we did here: we noticed a lack of concern for verifiability and a lack of self-efficacy (in the sense of knowledge and understanding).

Trust was pointed out by Schneider et al. in \cite{DBLP:conf/re/SchneiderLCHSX11} as an important factor for participants, as people are aware of potential security issues. Here trust is also an identified mental model of voters.

\section{A mobile application for Selene} \label{sec:selene}

\subsection{Selene mechanism} \label{subsec:selenemech}

\subsubsection{Protocol overview} Most verifiable voting schemes involve voters seeing and handling cryptographic data which can lead to errors or misuses \cite{DBLP:conf/uss/AcemyanKBW14}. Selene \cite{DBLP:conf/fc/RyanRI16} is an e-voting protocol that has been designed to provide an easier and more intuitive verification procedure for voters. It lets the voters verify that their votes have been included in the tally using a unique tracking number. To protect against coercion threats, i.e. achieve receipt-freeness and coercion mitigation, voters first learn their private tracking number after the votes have been posted. 
Selene uses ElGamal encryption, that is homomorphic and can act as a commitment scheme. An ElGamal encryption is a pair $(\alpha, \beta)$. For a given voter, the tracking number is encrypted using ElGamal and the $\beta$-term is published at the beginning of the election. The $\alpha$-term is kept secret and shared between several entities called Tellers. After the tally has been published, the $\alpha$-terms are sent to the voter, who can decrypt the tracking number with her key.
Full details about the cryptographic mechanisms can be found in the original paper \cite{DBLP:conf/fc/RyanRI16}.

\subsubsection{Voter experience}
As in Ryan et al. \cite{DBLP:conf/fc/RyanRI16} we assume that the voter already has the cryptographic key material needed for the protocol, i.e. we skip the key setup phase. The voting ceremony without coercion is as follows:\\
(1) The voter receives an invitation to vote.\\
(2) The voter makes her vote choice in the provided application, encrypts and signs the vote, and sends it to the Election Server.\\
(3) \textit{(Optional)} The voter later receives an invitation to visit the bulletin board when votes and tracking numbers are published.\\
(4) The voter receives the $\alpha$-term, and can retrieve the tracking number to verify her vote.

The third step is optional as it is only necessary if the voter is being coerced. For our implementation, we will assume that no coercion is happening and thus 
the third step is not available. Moreover, we simplified the fourth step by automating the $\alpha$ retrieval and tracker computation. The detailed methodology deployed during the user tests is described in section \ref{sec:protocol}.

\subsection{Application Design}

\subsubsection{A user-oriented approach} 

\begin{figure}[ht]
\centering \includegraphics[width=10cm]{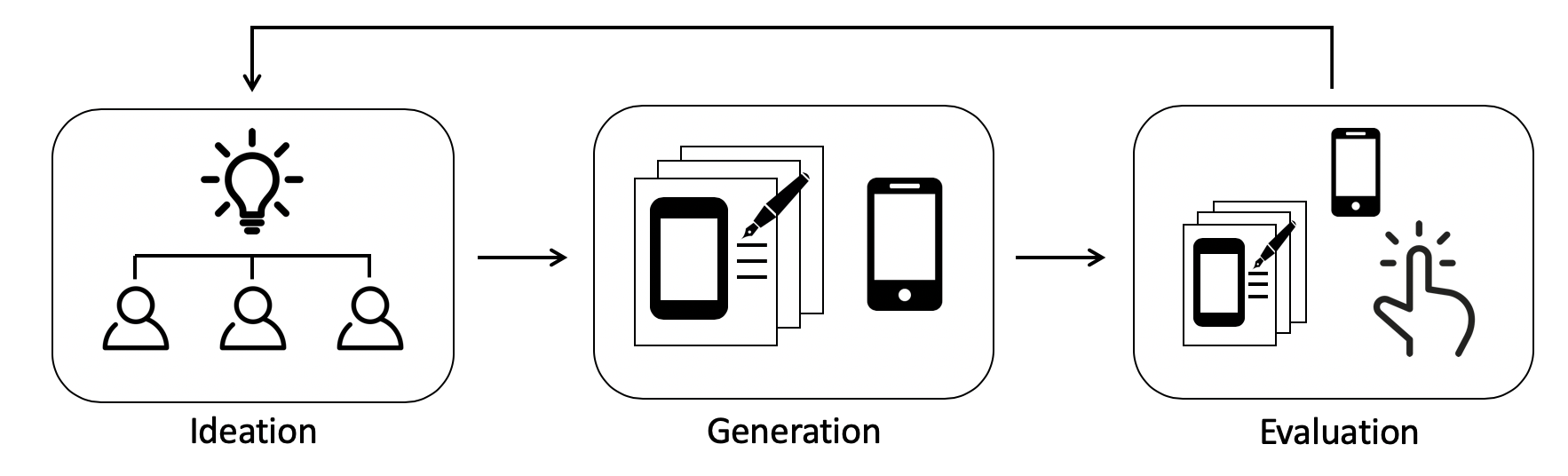}
\caption{Iterative process\label{fig:it}.}
\end{figure}

We followed a user-centred design methodology, which has originally been described by Norman \cite{Norman:1986:UCS:576915} and then detailed as a design process \cite{cooper2014face,normaneveryday}. In particular, we followed the UXD by Lallemand et al. \cite{LallemandCarine2018MddU}. The process consists of five steps which are \textit{planning}, \textit{exploration}, followed by an iterative process (shown in figure \ref{fig:it}) with \textit{ideation}, \textit{generation}, \textit{evaluation}.

\vspace{-.3cm}
The exploration phase includes a collection of user needs, and can be done using various methods, such as the a literature review of previous studies, interviews, focus groups or observations. In our case, we discussed the voting issues mentioned in several papers \cite{DBLP:conf/uss/AcemyanKBW14,DBLP:conf/hci/AcemyanKBW15,DBLP:conf/uss/GreeneBE06,DBLP:conf/uss/HerrnsonNHBCT06,DBLP:conf/stast/KarayumakKOVV11,DBLP:conf/chi/MarkyKRV18} during meetings with Human-Computer Interaction (HCI) experts who helped us develop and test prototypes of the e-voting application in a user-centred process in close collaboration.

Then we focused on the iterative process: we worked together with the HCI experts for the generation of ideas for the design during group sessions with up to ten group members. 
We then came up with the concept for a mobile application, that will have both features of voting and verification. We developed a first version that is a low-fidelity paper prototype. We evaluated this version with HCI experts. The received feedback on design and understanding of security allowed us to iterate and develop a second paper prototype, that was tested with both HCI and security experts.
The final iteration was a high-fidelity software version that will be described in detail in subsection \ref{subsec:impl}.

A particular challenge for the user experience of Selene is that a certain level of understanding might be necessary to achieve fulfillment of privacy needs: as secure and easy-to-use as the application might be, displaying the plaintext cast vote to the voter after election could seem insecure.
Further, the verification phase is not commonly used in standard elections and is largely unknown to users. Following a UXD process, we tried to anticipate users' expectations on voting and their questions on such a protocol, hence we designed an interface that, hopefully, is more understandable.
\vspace{-.3cm}

\subsubsection{Cryptography}

Selene uses several security mechanisms, however, the cryptographic details can be hidden from the user during the voting and verification phases. As mentioned above, we assume here that the voter doesn't have to configure her device with her secret keys or explicitly handle other keys such as the election public key. 

As mentioned before, the tracking number retrieval (fourth step in the voting experience) is simplified here and the voter simply has to click on a single button to use the $\alpha$-term, to download the $\beta$-term and to compute the tracking number. It will highlight the result on the bulletin board, displayed in-app, automatically. The voter's trapdoor key won't be explicitly manipulated by the voter and it is embedded in the phone, unlocked by the voter's credentials.

The other primitives used in Selene do not require direct interaction with voters (e.g. zero-knowledge proofs, mixnet, PET tests). Hence these are not mentioned in the conducted user experience test. In a real election implementation all of this can be public and verifiable by observers and interested voters.

We emphasise that this implementation is a first step that provides a user interface, in order to answer our research questions on user experience. This application is not ready to be used in a real election as both software security and the full cryptographic features have not been integrated yet.
As described in the protocol, the public key, the encrypted tracker, the commitment and the encrypted vote should be displayed on the bulletin board after every vote update. In this study we simplify and only update the bulletin board in the end.
\vspace{-.3cm}
\subsubsection{Trust assumptions}
Even if not all of the cryptographic primitives have been integrated in this version, we can already discuss the consequences of the design choices on the security properties. Firstly, we assume that the voting device is trusted for privacy. Further, in this test we have used a single device for voting and verification. In real scenarios, we would recommend that different devices, or at least apps, are used for vote-casting and vote-verifying for improved security.

The reason for using only one device was to simplify the experience for the participants focusing on a basic voting and verifying experience and to test this. The tracking number retrieval is also automated: the voter does not have to manually combine the $\alpha$ and $\beta$ terms and decrypt the tracking number. 
Since the $\alpha$ term is not shown to the voter, no visible $\alpha$ term needs to be faked, but a coercion-mitigation mechanism stills needs to implemented in the app to fake the tracking number itself. Further, the level of receipt-freeness in Selene will also depend on the chosen vote-casting method, e.g. a Helios type of electronic ballot will only achieve software-dependent receipt-freeness.  
However, this is a first iteration in the UX development of Selene, and the feedback from the participants given in section \ref{sec:mm} will help us to take the correct direction in the future developments. 

Finally, the verification phase was mandatory as a part of the test procedure. 
But in a real election it is to be expected that not all of the voters will verify their vote. We have not investigated the voters' motivation yet.

\subsection{Interface implementation} \label{subsec:impl}

The final application has been developed with the Android native language (Java) and the back-end server is developed in php and deployed on an Apache server. No security analysis has been performed as the goal of this interface is to run user tests.
The security of the application remains basic.
We describe below the interfaces provided.

\subsubsection{Android application}
The final application contains the two phases: one for voting and one for verification. The application retrieves flags related to the voter after authentication, that indicates: the voter's state (has voted (\texttt{true/false})), and the election's state (\texttt{vote/verify}).
\vspace{-0.3cm}

\begin{figure}[ht]
\centering \includegraphics[width=11cm]{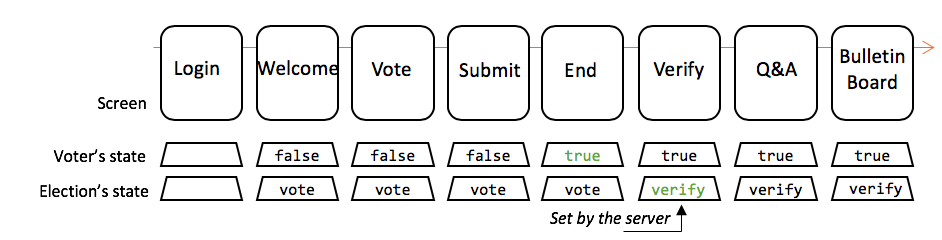}
\caption{Application workflow with states\label{fig:workflow}.}
\end{figure}
\vspace{-.3cm}
Figure \ref{fig:workflow} shows the organization of the application with the corresponding flags.
The application has been developed with a linear workflow, the voter only has a minimal choice for navigation, namely going backwards or forwards.
One should notice that in the context of a real election, the application might contain more screens with additional information. However, we will discuss in section \ref{sec:discussion} the advantages of such a linear construction.

\subsubsection{Administration page}
The back-end server is used to verify voter eligibility during authentication, to receive votes sent by the app, and tally the results.
When tallying the results, all pairs (tracker, vote) are counted and published on the Bulletin Board.
When allowing a voter to verify, the flag for the election's state is set to \texttt{verify} and the voter will be able to go through the verification workflow in the application.

\subsubsection{Bulletin Board}

The bulletin board is retrieved during the verification phase by the application. An additional button lets the voter highlight her vote. For this experiment, the bulletin board was accessible on the phone only but can be accessed directly from any browser, however it only contains minimal data needed for the user test.

\section{User Testing protocol} \label{sec:protocol}

\subsubsection{Participants}

We recruited 38 French participants (19 male and 19 female) through social networks, trying to ensure a fair distribution of our sample in terms of gender, age and education level. The average age was 35,4 years old (Min=19, Max=73, SD=12,45).
The education level broadly varied as well: no diploma (13\%), A-Levels (29\%), some college degree (21\%), Bachelor (18\%), Master (16\%) and PhD (3\%). The study has been run in French and the data presented has thus been translated into English.

To make their answers consistent and accurate, we selected participants that had participated at least in one political national election in France. 

\subsubsection{Procedure}

We provided each participant with a paper sheet explaining the context of the user test, that is a national election in France, together with the candidates' programs. Two personalized letters were distributed to each participant to provide them their individual credentials to access the application. Then the sessions were split up into 4 phases: (1) the voting phase, (2) a semi-structured interview, (3) the verification phase and (4) a semi-structured interview\footnote{A questionnaire about User Experience and Psychological Needs Fulfilment were also filled by participants during phases (2) and (4). The analysis is discussed in an other paper \cite{verena}.}. Before the verification phase, we gave them a second letter which was an invitation to verify their vote using the application.

\subsubsection{Methodology} 

The goal of the present analysis is to identify which mental models participants have of privacy and verifiability in e-voting.
The semi-structured interviews entailed the following topics: general opinion about the application, trust, control, understanding of the verification phase and of the bulletin board.
The three first topics were addressed after both the voting and verification phases. The two last topics were addressed after the verification phase only. 
We avoided security priming by not addressing security-related topics (such as privacy) until the very end of the study in order to avoid influencing participants' answers. In most cases, they mentioned by themselves the different security issues they could face with regards to e-voting. We describe in section \ref{sec:mm} which mental models we identified.
Information about the verification procedure was provided through paper letters and inside the application. 
The Q\&A screens are mandatory in the workflow and the participants have to go through them before verification. We told the participants that the tracking number let them verify that their vote has been counted in the final tally, that it helps to validate the election results, that this tracking number is unique and that the count can be verified by anyone. As we did not want to prime participant with possible security issues, we have not mentioned the risks of using one device and the associated trust assumptions.

\subsubsection{Data analysis} 

The user test was devised as a between-subjects study, and two versions of the e-voting application have been tested with our participants: half of the participants tested a baseline version and the other half an extended version where security aspects are additionally displayed to the participant. It contains additional information about the ongoing process in the application yet with no extra interaction. The impact of displaying security-related mechanisms was the topic a recently published paper \cite{verena} alongside with additional factors impacting UX (attractiveness, novelty, etc.) and psychological needs (autonomy, competence, etc.). Interestingly, we noted that this additional layer of communication remained largely unseen by the participants, with the perceived security being rated as only slightly higher in the elaborated version. The analysis of this paper focuses on interviews only and explore the feedback of participants regarding the security properties of voting, to check their understanding as it will be described in the next section.

To analyze the data retrieved from the interviews, we followed the methodology described in \cite{DBLP:books/el/LFH2017}. 
We coded the data through a theoretical thematic analysis, to look for patterns relating to the voting security properties. We organize the participants' answers into a list of concepts. We classified these concepts in categories given in section \ref{sec:mm} to understand voters' mental models of security. The categories were organised to match with the known theoretical models of security of e-voting: privacy properties including ballot-secrecy and coercion-resistance, and verifiability.

The qualitative analysis of answers in the semi-structured interviews leads to similar concepts for both versions, and we will thus analyze the participants' feedback in a similar way without considering the tested version.

One goal of user-centred design is to achieve a better alignment between participants' mental models and researchers' security vision, by "ensuring that products do fit real needs, that they are usable and understandable" \cite{normaneveryday}, we will discuss this in section \ref{sec:discussion}.

\subsubsection{Ethics}

The study follows the guidelines provided by the ethics commission at the author’s institution and was conform to GDPR.

\section{Mental models} \label{sec:mm}

\begin{figure}[ht]
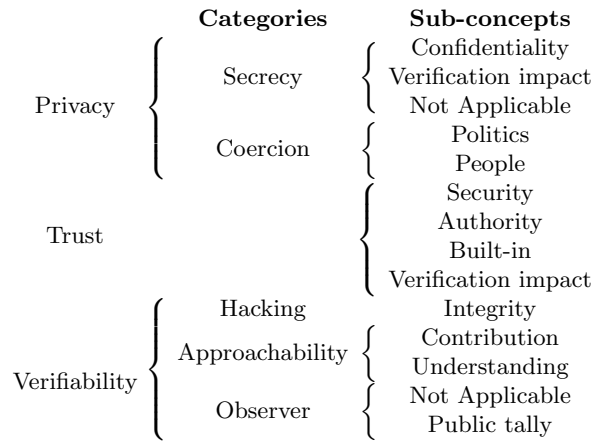

	\centering
	\begin{tabular}{ccccc}
       & & \textbf{Categories} & & \textbf{Sub-concepts}\\
      \multirow{5}{*}{Privacy}& \multirow{5}{*}{\hbox{$\left\{\vbox to 30pt{}\right.$}} & \multirow{3}{*}{Secrecy} & \multirow{3}{*}{\Bigg\{} & Confidentiality\\
      & & & & Verification impact\\
      & & & & Not Applicable\\
      & & \multirow{2}{*}{Coercion} & \multirow{2}{*}{\bigg\{} & Politics\\
      & & & & People\\
      \multirow{4}{*}{Trust} & &  & \multirow{4}{*}{\hbox{$\left\{\vbox to 25pt{}\right.$}} & Security \\
      & & & & Authority \\
      & & & & Built-in \\
      & & & & Verification impact \\
      \multirow{6}{*}{Verifiability} & \multirow{5}{*}{\hbox{$\left\{\vbox to 30pt{}\right.$}} & Hacking & & Integrity \\
      & & \multirow{2}{*}{Approachability} & \multirow{2}{*}{\bigg\{} & Contribution \\
      & & & & Understanding\\
      & & \multirow{2}{*}{Observer} & \multirow{2}{*}{\bigg\{} & Not Applicable \\
      & & & & Public tally
    \end{tabular}
    \caption{Mental model categorisation \label{fig:mms}}
\end{figure}

In \cite{Norman:1987:OMM:58076.58097}, Norman defined mental models as being "people's views of the world, of themselves, of their own capabilities, and of the tasks that they are asked to perform, or topics they are asked to learn". The interactions they have with the environment make them form internal models of the system they are interacting with. 
We here propose a categorization of participants' feedback. An overview is given in figure \ref{fig:mms}. From the identified concepts, we derive a categorization of the mental models expressed by participants. We will discuss how these results should impact the future development of the application in section \ref{sec:discussion}.
\vspace{-0.3cm}

\vspace{-0.6cm}
%Naturally, there are overlaps between the topics. 
%Trust concepts, especially security, are related to the privacy and hacking concepts. The approachability concepts are related to the Secrecy model, but also to integrity of votes and understanding of the application.

We now explain this structure in detail.

\subsection{Privacy}

\subsubsection{Secrecy Mental Model}
Participants mentioned that their vote must be kept \textit{confidential and anonymous}. They questioned the data management, wondering if someone knows the relation between identities and votes. Some participants stressed the importance of the booth, another argued that the booth is not private either, "some people are looking" (P10).

The \textit{verification} could have a negative impact on secrecy of votes as well, "it is like someone else could see it too" (P22).

Finally, another concept was \textit{not applicable}, as some people did not feel concerned by secrecy, as "others know who I am voting for" (P13) or "you have to take responsibility for your political decisions" (P38).
\subsubsection{Coercion Mental Model}
Coercion from \textit{people} (in the sense of a physical attempt to coerce) was mentioned several times. In particular, participants mentioned the advantage of being able to vote at home, as other people won't influence them: "Here I don't have interactions with other people" (P5), or "I am sure to make my own choice [...] I feel less pressure than in polling station, with people behind" (P2). Vote buying was mentioned once "We can be manipulated, one could buy our vote, but we need to evolve" (P23).

Finally, the \textit{political aspect} of coercion was also mentioned a few times, as some parties could try to cheat and to steal credentials from voters: "We must pay attention to parties, ensure there is no violation, that the elderly or other vulnerable persons do not get their vote stolen" (P10).

\subsection{Trust}
The concept that appears the most for Trust is \textit{security}. Participants mentioned that their trust in the application is highly dependent on the security provided: "I don't trust it, how could we know if it is really secure?" (P29), "I trust it, there are breaches everywhere but I think we can secure this" (P10). In particular it was reflected on their other mental models related to privacy and efficiency. Hence, we can derive this security concept with the following sub-concepts: coercion, secrecy, and understanding that increased or decreased the security perceived. 

Another concept was \textit{authority}, mentioned as trust-transference in \cite{DBLP:journals/corr/AliM16}. Participants refer to some trusted third party to emphasize their own trust: "if it is done by an authority, I will trust it" (P2) or "I trust the government, they will do what is necessary to ensure vote security" (P12). They rejected the verification process arguing with their trust in the authority: "If I trust the application I don't see why we should verify that the vote has been taken into account" (P12).

Some participants expressed a \textit{built-in} trust, e.g. "I always trust technology" (P14) or "I trust it as I would trust any mobile applications".

A \textit{verification impact} was raised, mostly decreasing trust, e.g. "I don't trust the application after verification, even if the tracking number is private" (P33), even though an opposite positive effect on trust was also mentioned by some users: "the second phase makes me feel secure" (P4).

\subsection{Verifiability}

\subsubsection{Hacking Mental Model}
Participants were concerned by the security of internet technologies and had many preconceptions. Even if participants didn't master the complexity of internet security, they were aware that it could be an issue. For example, they mentioned problems they heard about other voting systems with electronic ballots: "In United States there was this elections hacking. Paper is more reliable" (P15). Others feared internet technologies in general: "I think internet is vulnerable, even if the app is secure" (P24). Ballot stuffing was also mentioned as a big problem: "There are people who can buy hackers' services to have thousand of votes added, we will never know." (P28).

\textit{Integrity} is a concept that often appeared during the interviews. Participants questioned the good behavior of the application as they did not receive any proof of it. The reliability of the system is questioned: "It does not guarantee that it is really my vote." (P33). Some participants also expressed the need for a procedure in case of encountering an issue: "Who should I call in case of problem? And if my vote is not in the list?" (P19) or "If I voted A and it shows B, what should I do?" (P32).

\subsubsection{Approachability Mental Model}

Some participants were convinced of the good behavior of the system due to the verification phase. It was mentioned as a proof of their personal \textit{contribution} to the elections: "Seeing that my vote is taken into account, seeing others' votes, it lets me believe that I contribute to something" (P18) or "It is important to see that my vote has been counted" (P27).

Most participants understood that they were seeing a confirmation of their tallied-as-intended vote. But they expressed their [lack of] \textit{understanding} in the process of verification: "I feel in control maybe because I can see what I did, I can see my vote again" (P11) or on the contrary "I wonder why this is here, seeing results with percentages is enough for me" (P3).

We tried to rate participants' understanding of Selene's mechanism, through the two last questions stated in section \ref{sec:protocol}. To help participants to answer, we provided some light explanation of the verification phase meaning. However, many participants did not manage to provide a complete description of the verification phase after using the app. 
Furthermore, the tracking number has not always been understood as such, but rather as a counting of votes: "We can see our candidate and the number of people who voted for him" (P6).

\subsubsection{Observer Mental Model}
Some participants stressed the importance of observation. In France, voters are allowed to go to the polling-station to observe the public count of votes. However, most of our participants did not noticed the link between this real-life procedure and the availability offered by the bulletin board: "The list is not really informative" (P35), "I can't see if there is any interest to see this list with all details" (P17). This can be explained by their lack of understanding of the procedure, as compared to a physical count of votes, in which they can see and understand each step: "In polling stations you can verify by yourself, on internet it's questionable" (P24).

Finally, the \textit{individual} aspect seems to be enough to participants, e.g. "Seeing percentages with general results, and my individual vote is enough" (P17).

\section{Discussion} \label{sec:discussion}

Norman in \cite{Norman:1987:OMM:58076.58097}, and Cooper et al. in \cite{cooper2014face}, show that three models must be considered in the design of a user interface: the system or implementation model that reflects how the system actually works, the system image or represented model that reflects what is shown to the user and the mental model that is the projection made by the user. Here we focus on the discrepancies between those three distinct categories.
The goal of a user-oriented design process, such as the UXD process, is to provide an interface, a represented model, that is close to the user's mental model and that remains accurate with the system model.

\subsection{Comparison between mental models and security properties}

The properties on which we base the implementation model of a voting scheme are Privacy properties and Verifiability. Selene provides ballot-secrecy, receipt-freeness and has a coercion mitigation mechanism. It also provides individual and universal verifiability. However, as mentioned in the previous subsection, the coercion mitigation mechanism has not been implemented.

Despite this, voters were concerned about Coercion and about Privacy during elections in general. Mental models for Privacy were consistent with the properties of the system, and the reason might be that
Privacy is a mandatory element required by law during elections in France, and it is taught to people at an early age at school. 

On the other side, the novelty of the verification phase seemed to prevent participants from properly explaining their experience.
However, indirect properties and potential issues were mentioned, such as hacking and integrity, and public tallying. It appears that participants were able to point out the potential issues of online voting without seeing that the verification mechanism was part of the solution.

Also for privacy, we can argue that an early education on verifiability could lead to a better understanding and acceptance of the concept. 

Trust is not a security property of voting protocol. However, it plays an important role for voters and impacts the use of a system. This aspect is important for people to accept the system they use.

Even if the convenience of online voting was mentioned many times, voters stressed their lack of knowledge about internet technologies as a big drawback. Paper-ballot voting contains steps that are understandable and accessible to people, and this is not in general the case for online voting. Even if this aspect is not required by law as in Germany, it seems reasonable that voters are more willing to trust a process they fully understand.

\subsection{Impact of a user-centred application on the voting experience}

First of all, we observed 100\% of effectiveness for vote casting: all participants were able to cast their vote successfully.\footnote{The verification phase was mandatory in our experiment and everyone managed to go through the verification workflow. But not all participants understood what was happening and we can't ensure that the effectiveness would be as high for verification if it is not mandatory.} The application was designed in order to be easy-to-use and responding to users' expectations, and we can argue that it is the linearity of the vote casting in the Selene implementation that leads to this excellent result. 
The quality of this straightforward behaviour was mentioned several times by participants, e.g. "we follow the workflow but we can't really make a mistake" (P3).

Another explanation for the observed usability can be the design of the protocol itself. As mentioned in the introduction, Selene was designed in order to reduce complicated interactions with users, and to be more intuitive. Helios is another e-voting scheme that requires, or at least suggests, voters to perform audits of their ballots through a Benaloh's challenge \cite{DBLP:conf/uss/Adida08}. This often leads to a lower effectiveness rate: a study from Marky et al. \cite{DBLP:conf/chi/MarkyKRV18} has shown that this procedure is considered as counter-intuitive by participants. Indeed, participants who audited their ballot did not understand why they were not allowed to cast the audited ballot after-all. This kind of step does not occur with Selene, as the verification happens after the end of the election. The voting phase is thus not burned with a verification step. Moreover, the authors of \cite{DBLP:conf/chi/MarkyKRV18} showed that automation of the verification feature improved effectiveness. In our application, we automated the retrieval of the tracking number. Instead of asking participants to manipulate $alpha$ and $beta$ terms, we retrieve them and computed the tracking number automatically.

The questionnaires analyzed in \cite{verena} showed that the usability aspects, i.e. efficiency, perspicuity and dependability \cite{ueqhandbook}, scored above average. 
Despite this and the fact that all voters managed to cast their votes% effectiveness rate of 100\%
, the participants' feedback show that our application needs more development iterations to be better consistent with voters' model of voting.

Also, simply displaying information about security features  to the participants was not enough to make them explicitly see it (as discussed in \cite{verena}). However, we have seen here in our analysis that the security of such an application is an important factor for trust and it was emphasized for the secrecy and verifiability concepts. Moreover, the information related to security in the expanded version of the app was shown during loading screens. 
It might be that the progress bar prevented the participants from reading the information displayed below\footnote{An other study could verify where the information displayed would be more visible.}. In a small study \cite{DBLP:journals/ieeesp/KulykNBV17}, the authors found that the voters chose more secure systems as their preferred scheme even if they scored lower on the SUS scale. 
It is thus interesting whether allowing more cryptographic interactions could increase the acceptance, even if it reduces the usability. One idea could here be to also implement coercion mitigation mechanism. This mechanism allows the voter to ask for a fake tracker in case of coercion. It might be that voters have missed this mechanism to understand the verification phase. 

Indeed, the meaning of the verification phase and in particular of the tracking number was explained through Q\&A screens. However, many participants did not understand fully, or were not able to describe the verification phase.
As Acemyan et. al observed in their study \cite{DBLP:conf/hci/AcemyanKBW15}, when participants were requested to draw their mental model, they expressed the voting steps for each tested scheme and avoided the verification steps and the verification phases of each system was considered useless in many cases, like in our observations.
On the other hand, participants who understood the interest of seeing their vote in the app did not understand why they were seeing others' votes, as their own vote and only this vote was highlighted in the application.
The implementation of the coercion mitigation mechanism could also help here, however this assumption needs to be tested in a new iteration of the application.

Olembo et al. \cite{olembo/motivation} showed that specific messages could motivate voters to verify their vote, as they understand better the objective of such a procedure. In particular, they focused on risks, norms and analogies. In our application, the focus has been done on norms only, i.e. we explained what is the purpose of verification and what it brings to society. We emphasized democracy protection and integrity of votes records. Now, according to voters' models for Coercion and their concerns on hacking, a stronger emphasis on the incurred risks and solutions provided by verification might help the voter to understand.
Some people understood that the tracking number was instead the number of people voting like they did. A simple improvement is to add letters to the tracking number. 

In this version of the application, the bulletin board was not accessible before the individual verification phase\footnote{See third phase of the voter experience described in subsection \ref{subsec:selenemech}}. One improvement could be to make the bulletin board available before the individual verification. The possibility to request a fake tracker must also be implemented at this stage. 
In addition, once the Selene check is doable, we could show the individual vote first (with fake or real tracking number) and let the voter consult the bulletin board on purpose.
\vspace{-.5cm}
\subsubsection{Limitations}

The results of our study are bounded by some limitations. \\
First, the user tests were done in a laboratory that had a reassuring impact on participants. Some of them admitted that they were not really feeling any threats for their vote as they were part of an experiment. The influence in a lab context on user studies is discussed in \cite{JUSLallemand}.\\
We mention to our participants that the elections were related to the national elections in France, however we did not use real candidate names nor run an election that already happened, as suggested in \cite{markyzollinger}.\\
Also, the participants had a very limited amount of time to understand the verification procedure, and the novelty of such a protocol might require more time to be understood and accepted. A broader context would be provided in real elections, giving users time to understand the process of verifiability of the application.
We also assumed in our study that the configuration of the devices was already done. 
The ease of use might be questioned if the registration to online voting and keys configuration must be performed by voters.
However, this configuration could be done only once for several elections. \\
Finally, we asked the participants to verify their vote right after the vote casting phase. In other protocols and user studies, the verification is done during the vote casting (e.g. Benaloh's challenge, or return codes). In this protocol, the verification is performed after the results have been published and due to experimental constraints the participants had to do it right after vote casting, that could have disconcerted them. 

\section{Conclusion and Future Work} \label{sec:ccl}

In this paper, we have provided the first interface prototype for the Selene e-voting protocol. We have followed a user-centered design, UXD,  where the usability is enhanced and cryptographic interactions have been hidden. This approach has consequences on trust assumptions for the voting protocol, but has provided insights on the mental models of Privacy and Verifiability.
User tests have highlighted possible improvements on our application for Selene, but it has also raised more general concerns we need to consider in the design of e-voting protocols.
We have seen that mental models for Privacy with Secrecy and Coercion were consistent with the voting protocol concepts.
However, the understanding of the verification phase has to be facilitated. We have seen that the lack of understanding could lead to trust issues: participants questioned integrity of the elections and the purpose of the verification phase.
Voting schemes are developed today to be end-to-end verifiable, but verification is not natural to users and voters need more time to accept it and understand it. An easy-to-perform mechanism for verification like the one described in Selene has been effective but is not enough to convince voters of the security behind the scheme. 
For future work, the implementation of missing mechanisms for Selene must be performed in order to provide a complete experience to voters. A new iteration of the application (using two devices) based on the received feedback is being developed in order to increase the understanding of voters, and reassure them of the security mechanisms in use.

\subsubsection*{Acknowledgements}
We would like to thank the Luxembourg National Research Fund (FNR) for funding, in particular PBR was supported by the FNR INTER-Sequoia project which is joint with the ANR project SEQUOIA ANR-14-CE28-0030-01, MLZ was supported by the INTER-SeVoTe project and VD was supported by FNR grant number PRIDE15/10621687.

\bibliographystyle{splncs04}
\bibliography{main}

\end{document}